# Accelerating *GW* calculations of point defects with the defect-patched screening approximation

Du Li[‡], Zhen-Fei Liu*[,§], and Li Yang[†, ‡, ∥]

[‡]Department of Physics, Washington University in St. Louis, St. Louis, Missouri 63130, USA

[§]Department of Chemistry, Wayne State University, Detroit, Michigan 48202, USA

[∥]Institute of Materials Science and Engineering, Washington University in St. Louis, St. Louis, Missouri 63130, USA

**ABSTRACT:** The *GW* approximation has been widely accepted as an *ab initio* tool for calculating defect levels with many-electron effect included. However, the *GW* simulation cost increases dramatically with the system size, and, unfortunately, large supercells are often required to model low-density defects that are experimentally relevant. In this work, we propose to accelerate *GW* calculations of point defects by reducing the simulation cost of the many-electron screening, which is the primary computational bottleneck. The random-phase approximation of many-electron screening is divided into two parts: one is the intrinsic screening, calculated using a unit cell of pristine structures, and the other is the defect-induced screening, calculated using the supercell within a small energy window. Depending on specific defects, one may only need to consider the intrinsic screening or include the defect contribution. This approach avoids the summation of many conductions states of supercells and significantly reduces the simulation time. We have applied it to calculating various point defects, including neutral and charged defects in two-dimensional and bulk systems with small or large bandgaps. The results consist with those from the direct *GW* simulations, and the agreements are further improved at the dilute-defect limit, which is experimentally relevant but extremely challenging for direct *GW* simulations. This defect-patched screening approach not only clarifies the roles of defects in many-electron screening but also paves the way to fast screen defect structures/materials for novel applications, including single-photon sources, quantum qubits, and quantum sensors.

## 1. Introduction

Point defects decide a wide range of fundamental materials properties, such as transport, optical, and magnetic properties. For example, certain defects are responsible for features in optical spectra and hence the name of "color centers", such as ultraviolet color centers in zinc oxide and cubic boron nitride [1-5], visible and near-infrared color centers in diamond and silicon carbide [6-19]. In recent years, there has been growing interest in applying materials with point defects in single-photon sources, quantum qubits, and quantum sensors [20,21]. For instance, in diamond and silicon carbide, point defects not only act as quantum emitters and sensors but also possess spins that can be used to implement isolated quantum qubits [14-18,22,23]. Point defects in two-dimensional (2D) materials, such as chalcogen vacancies in 2D transition-metal dichalcogenides (TMDs) and point defects in layered hexagonal boron nitride (*h*-BN), have attracted significant attention for single-photon emitters [24-33] and quantum qubits [34-37]. These defects in 2D materials can have close interactions with the environments, making them promise for quantum sensing [38,39].

*Ab initio* electronic structure methods are indispensable in predicting microscopic structure-property relationships for defect systems. It is by now common knowledge that the local and semi-local exchange-correlation functionals within the framework of Kohn-Sham (KS) density functional theory (DFT) tend to underestimate the band gap of materials and may not provide accurate descriptions of the defect energy levels and how these levels are aligned with the band edges of the otherwise pristine material [40,41]. To overcome this limitation, the *GW* approximation within the framework of many-body perturbation theory (MBPT) [42,43] is a popular choice and has been shown to be quantitatively accurate for a broad range of defects [41,44-47]. However, when it is implemented using a plane-wave basis that is suitable for periodic systems, a large number of empty states need to be included in the calculation to converge the non-interacting KS polarizability, which is the computational bottleneck for calculating point defects in large supercells. To date, many methods have been proposed to make the calculations of large systems more tractable. The large number of empty states can be approximated by free-electron states [48] or be eliminated using the Sternheimer equation [49]. Moreover, calculations can be simplified by replacing the widely used plane-wave basis with others, such as Wannier orbitals [50-52], projected-dielectric eigen-decomposition basis [53-56], or stochastic orbitals [57,58], etc. However, these approaches require additional techniques to transform between different basis. In this work, we propose an accelerated *GW* approach in plane-wave basis, which is generally applicable to a broad range of defect systems.

We focus on overcoming the major computational bottleneck, namely the noninteracting KS polarizability within the random-phase approximation (RPA), in *GW* calculations of point defects. We present two-level approximations: intrinsic screening approximation and defect-patched screening approximation. At the first level, we approximate the RPA polarizability of a defect-containing supercell by that of the pristine structure, realized by the reciprocal-space folding [59]. This reduces the polarizability calculation to a unit cell, and the calculated defect levels excellently agree with the direct *GW* ones when the minimum energy gap between defect levels is close to the bulk band gap of the parent materials. Otherwise, if the defect gaps are substantially smaller than the bulk bandgap, we further include the defect-contributed RPA polarizability but only compute this contribution within a small energy window. As a result, this defect-patched approximation avoids including many unoccupied states to obtain the RPA polarizability, significantly reducing the computational cost. We demonstrate the strength of this approach in obtaining defect levels of a wide range of defect structures, including neutral and charged defects in 2D or bulk insulators and semiconductors. The agreements between the defect-patched approximation and direct *GW* result are excellent. They can be further improved at the dilute-defect limit, which is more relevant to experimental situations. Such an accelerated *GW* approach can serve as a powerful tool to screen point defects with desired defect levels for broad applications.

The structure of this article is as follows. In Section 2, we discuss the theoretical framework to calculate quasiparticle energies of point defect systems, showing the main assumptions to simplify the calculation of RPA polarizability. In Section 3, we demonstrate the strengths of our approach using various examples: the double carbon substitutional defect in *h*-BN in Section 3 A; the single selenium vacancy in monolayer $MoSe_2$ in Section 3 B; the single boron vacancy in monolayer *h*-BN in Section 3 C; the negatively charged NV center in diamond in Section 3 D. In Section 4, we provide discussions of our approach in two aspects: point defects in the dilute-defect limit and the comparison of computational efficiency. We conclude our work in Section 5.

**2. Theoretical Approach**

In the *GW* approximation, quasiparticle self-energy is calculated via $\Sigma = iGW$,[42,43] where $G$ is the Green's function and $W$ is the screened Coulomb interaction. The primary bottleneck of *ab initio GW* calculations is the high computational cost and memory requirement of the noninteracting RPA polarizability, $\chi^0$. In reciprocal space, the $\chi^0$ of insulators is:

$$\chi^0_{GG'}(q,\omega) = \Sigma_{v,c,k} \frac{\langle v, k+q|e^{i(q+G)\cdot r}|c,k\rangle\langle c,k|e^{-i(q+G')\cdot r'}|v,k+q\rangle}{\omega + E_{v,k+q} - E_{c,k}}, \qquad (1)$$

where $q$ and $k$ are vectors in the first Brillouin zone, $G$ is a reciprocal-space lattice vector. $v$ refers to occupied valence states and $c$ refers to unoccupied conduction states. $|v, k+q\rangle$, $|c, k\rangle$, $E_{v,k+q}$, and $E_{c,k}$ are the KS eigenvectors and eigenvalues, respectively. In plane-wave basis, Eq. (1) leads to a formal scaling of $O(N^4)$ or a practical scaling of $O(N^3 \log N)$ [60], where $N$ corresponds to the system size. To model isolated point defects or dilute defects of experimental relevance, a large supercell is needed, resulting in high computational cost.

To make the *GW* calculations more tractable, we approximate the RPA polarizability $\chi^0$ of a supercell with a point defect as:

$$\chi^0 \approx \chi^0_{\text{Intrinsic}} + \chi^0_{\text{Defect}}. \qquad (2)$$

$\chi^0_{\text{Intrinsic}}$ is the polarizability of the supercell without defect, which is essentially the intrinsic screening of the parent material. $\chi^0_{\text{Defect}}$ is the contribution from the point defect. In the following, we simplify the calculation of $\chi^0_{\text{Intrinsic}}$ and $\chi^0_{\text{Defect}}$, respectively.

$\chi^0_{\text{Intrinsic}}$ of the supercell is essentially the polarizability of pristine structures, which can be obtained by computing the $\chi^0$ from a pristine primitive cell followed by the reciprocal-space folding procedure [59]. This idea was originally developed in the context of weakly coupled molecule-substrate interfaces [59,61]. In this approach, the RPA polarizability of a supercell can be written as $\chi^0_{\text{Intrinsic}}(\boldsymbol{q} + \boldsymbol{G}, \boldsymbol{q} + \boldsymbol{G}')$, where $\boldsymbol{G}$, $\boldsymbol{G}'$ are reciprocal lattice vectors and $\boldsymbol{q}$ is a vector in the first Brillouin zone of the supercell. Similarly, the RPA polarizability of the primitive cell can be written as $\tilde{\chi}^0_{\text{Intrinsic}}(\tilde{\boldsymbol{q}} + \tilde{\boldsymbol{G}}, \tilde{\boldsymbol{q}} + \tilde{\boldsymbol{G}}')$, where $\tilde{\boldsymbol{G}}$, $\tilde{\boldsymbol{G}}'$ are the reciprocal lattice vectors and $\tilde{\boldsymbol{q}}$ is a vector in the first Brillouin zone of the primitive cell. $\chi^0_{\text{Intrinsic}}$ and $\tilde{\chi}^0_{\text{Intrinsic}}$ describe the same quantity, i.e., $\chi^0_{\text{Intrinsic}}(\boldsymbol{q} + \boldsymbol{G}, \boldsymbol{q} + \boldsymbol{G}') = \tilde{\chi}^0_{\text{Intrinsic}}(\tilde{\boldsymbol{q}} + \tilde{\boldsymbol{G}}, \tilde{\boldsymbol{q}} + \tilde{\boldsymbol{G}}')$, where $\boldsymbol{q} + \boldsymbol{G} = \tilde{\boldsymbol{q}} + \tilde{\boldsymbol{G}}$ and $\boldsymbol{q} + \boldsymbol{G}' = \tilde{\boldsymbol{q}} + \tilde{\boldsymbol{G}}'$ hold when these quantities are expressed in Cartesian coordinates. Therefore, the supercell form, $\chi^0_{\text{Intrinsic}}(\boldsymbol{G}, \boldsymbol{G}'; \boldsymbol{q})$, can be obtained from the primitive-cell form, $\tilde{\chi}^0_{\text{Intrinsic}}(\tilde{\boldsymbol{G}}, \tilde{\boldsymbol{G}}'; \tilde{\boldsymbol{q}})$.

It is challenging to directly calculate the second term ($\chi^0_{\text{Defect}}$) in Eq. (2). Fortunately, in diluted defect densities, the density of defect states is much smaller than that of the background crystal. This is also the typical experimental situation. Thus, $\chi^0_{\text{Defect}}$ can be regarded as a small correction. More quantitatively, according to the denominator of Eq. (1), the main contribution to polarizability is from the transitions between defect levels with small energy differences. Therefore, it is possible to calculate the defect contribution within an energy window. This idea was originally developed in the context of strongly coupled molecule-metal interfaces [62]. We propose to calculate the defect-contributed polarizability as:

$$\chi^0_{\text{Defect}}(\boldsymbol{q}, \omega) = \sum_{\substack{v,c,\boldsymbol{k} \\ E_{v,\boldsymbol{k}+\boldsymbol{q}}, E_{c,\boldsymbol{k}} \in [E_{\min}, E_{\max}]}} \frac{\langle v, \boldsymbol{k}+\boldsymbol{q}|e^{i(\boldsymbol{q}+\boldsymbol{G})\cdot\boldsymbol{r}}|c,\boldsymbol{k}\rangle\langle c,\boldsymbol{k}|e^{-i(\boldsymbol{q}+\boldsymbol{G}')\cdot\boldsymbol{r}'}|v,\boldsymbol{k}+\boldsymbol{q}\rangle}{\omega + E_{v,\boldsymbol{k}+\boldsymbol{q}} - E_{c,\boldsymbol{k}}}$$

$$- \sum_{\substack{v,c,\boldsymbol{k} \\ v,c \in \text{Intrinsic} \\ E_{v,\boldsymbol{k}+\boldsymbol{q}}, E_{c,\boldsymbol{k}} \in [E_{\min}, E_{\max}]}} \frac{\langle v, \boldsymbol{k}+\boldsymbol{q}|e^{i(\boldsymbol{q}+\boldsymbol{G})\cdot\boldsymbol{r}}|c,\boldsymbol{k}\rangle\langle c,\boldsymbol{k}|e^{-i(\boldsymbol{q}+\boldsymbol{G}')\cdot\boldsymbol{r}'}|v,\boldsymbol{k}+\boldsymbol{q}\rangle}{\omega + E_{v,\boldsymbol{k}+\boldsymbol{q}} - E_{c,\boldsymbol{k}}},$$

(3)

where the first term contains all transitions within the defined energy window $[E_{min}, E_{max}]$, including defect-defect, defect-bulk, and bulk-bulk contributions. The second term removes the non-defect contributions and removes the double counting. Although the defect polarization is calculated in supercells, it quickly converges as the energy window is increased. In all our calculated defect structures, the converged energy window only needs to extend into band edges around 1 eV. This defect-patched approximation avoids the bottleneck of the direct RPA polarizability calculation in Eq. (1), which requires a significant number of unoccupied bands/states of supercells.

In the following, we show that, depending on specific defects, one may only need to consider the intrinsic polarizability or include the defect contribution. These two tiers of approaches accelerate the *GW* calculations of defects and make it possible to accurately screen a large number of defect systems for targeted applications.

## 3. Applications

In this section, we demonstrate the accuracy of our approximations by comparing them with the direct *GW* calculations for several typical point defects of broad interests. Unless otherwise specified, the computational details of this section are as follows. The DFT calculation is performed using the Perdew-Burke-Ernzerhof (PBE) exchange-correlation functional [63] and a plane-wave energy cutoff of 60 Ry with the QUANTUM ESPRESSO package [64]. For two-dimensional (2D) structures, the vdW interaction is included via the semi-empirical Grimme-D3 scheme [65]. A vacuum of 18 Å between adjacent layers is used to avoid spurious interactions between periodic images along the out-of-plane direction for 2D structures. The *GW* calculations are performed using the BERKELEYGW package with our modifications of the calculating of polarizability [60] presented in Section 2. All *GW* calculations reported in this work are perturbative, i.e., $G_0W_0$. The Hybertsen-Louie generalized plasmon-pole model is used to treat the frequency dependence of the dielectric function [42]. The static remainder approximation [66] is used in the evaluation of the self-energy for faster convergence. The slab Coulomb truncation [67] is adopted to mimic suspended 2D structures.

**A. Double carbon substitutional defects in monolayer *h*-BN**

The first example is the double carbon substitutional defect ($C_BC_N$) in monolayer hexagonal boron nitride (*h*-BN), and it is proposed to be a candidate of single-photon emitters [68,69]. As shown in Fig. 1(a), this defect is formed by replacing two neighboring boron and nitrogen atoms with a pair of carbon atoms. We construct a 9 × 9 × 1 supercell with a point defect, corresponding to a 1.2% defect concentration. A 2 × 2 × 1 q-grid is adopted to calculate the polarizability and quasiparticle energies in the supercell. Around 8,000 empty states are included in the direct *GW* calculation of the RPA polarizability.

Fig. 1(b) presents the DFT electronic structure, where the gray lines are continuous bulk bands of *h*-BN, and the two blue lines are the defect energy levels lying inside the gap. To better illustrate how the defect levels are aligned in energy with the band edges of the *h*-BN, a schematic energy diagram is plotted in Fig. 1(c). As shown in Fig. (c), the minimum energy difference between occupied and unoccupied defect states is 3.58 eV, which is defined as the "defect gap" in this article. Meanwhile, the energy gap between continuum bulk states of *h*-BN is 4.68 eV, which we define as the "bulk bandgap" and is essentially the bandgap of pristine structures. The distance between the lower-energy defect level and the valance band maximum (VBM) of bulk states is 45 meV, and that between the higher-energy defect level and the conduction band minimum (CBM) is 65 meV. These results agree with previous DFT calculations [70].

Fig. 1(d) shows the quasiparticle defect levels of this 9 × 9 × 1 supercell with a $C_B C_N$ defect from a direct *GW* calculation. As expected, the reduced screening of such a 2D structure significantly enhances the self-energy correction. The bulk bandgap is increased from 4.68 eV to 7.29 eV, and the defect gap is increased from 3.58 eV to 6.38 eV.

We first neglect defect contribution and only consider the intrinsic screening in the *GW* calculation [$\chi^0 \approx \chi^0_{\text{Intrinsic}}$, i.e., only including the first term in Eq. (2)]. This will be called "intrinsic screening approximation" in this paper. The results are presented in Fig. 1(e). The quasiparticle bulk bandgap of *h*-BN is 7.32 eV, and the defect gap is 6.43 eV. Compared to the direct *GW* results, the agreements are surprisingly good, and the difference is within 50 meV.

Then we further consider the "defect-patched screening approximation" by adding the defect contribution to the polarizability, i.e., including both terms in Eq. (2) with the second term computed via Eq. (3). We find that the result converges within a relatively narrow energy window. For this defect, the energy window of summation is set to be 1.8 eV below the VBM and 1.7 eV above the CBM of bulk h-BN, as marked by the black-dash lines in Fig. 1 (b). The *GW* results are plotted in Fig. 1(f), in which the bulk bandgap of *h*-BN is 7.31 eV and the defect gap is 6.41 eV. The difference in the defect gap between the direct *GW* and defect-patched screening approximation is slightly improved to be within 30 meV, compared with the intrinsic screening approximation [Fig. 1(e)].

The success of intrinsic screening approximation and the small defect contribution to the polarizability are expected, and the reason can be seen from Eq. (3). The polarizability is affected by transition energy, which is strongly affected by defect gap. For the $C_BC_N$ defect, the defect gap is large (comparable to the bulk bandgap), resulting in a large dominator in Eq. (3). As a result, the defect contribution is negligible, making the intrinsic screening approximation reliable.

This explanation can be further justified by comparing the RPA polarizability matrices in plane-wave basis. The matrix elements of the polarizability of the direct *GW* calculation are plotted in Fig. 2(a), and the difference from the intrinsic polarizability ($\chi^0 - \chi^0_{\text{Intrinsic}}$) is plotted in Fig. 2(b). We can easily find that the difference is about two orders of magnitude smaller than that of the directly-calculated $\chi^0$, numerically justifying the intrinsic screening approximation. We have further plotted the matrix elements of $\chi^0_{\text{Defect}}$ in Fig. 2(b). Their magnitude is also about two orders of smaller than those of the directly-calculated $\chi^0$, supporting that the defect contribution to the polarizability is negligible.

Finally, considering the experimentally relevant diluted defect densities, where the density of defect states will be much lower than that in the $9 \times 9 \times 1$ supercell. As a result, the defect contribution to the total polarizability will be even smaller for diluted defect densities, further supporting the intrinsic screening approximation. Therefore, we expect that the intrinsic screening approximation is reliable for defect structures with a large defect gap.

**B. Single selenium vacancy in monolayer MoSe$_2$**

In this section, we investigate a selenium vacancy ($V_{Se}$) in monolayer MoSe$_2$, which is formed by removing a single selenium atom in monolayer MoSe$_2$ as illustrated in Fig. 3(a). $V_{Se}$ in monolayer MoSe$_2$ and many other similar point defects in layered TMD are important structures that can be used to realize 2D single-photon emitters [24-28]. In our simulation, a $5 \times 5 \times 1$ MoSe$_2$ supercell is constructed corresponding to a 2% vacancy concentration. A q-grid of $3 \times 3 \times 1$ is adopted to calculate the dielectric function and quasiparticle energies in the supercell. Around 4,300 empty states are included in the direct calculation of the RPA polarizability.

The DFT band structure and corresponding schematic energy diagram are shown in Fig. 3(b) and Fig. 3(c), respectively. The DFT bulk bandgap of monolayer $MoSe_2$ is 1.43 eV. There are nearly double-degenerate unoccupied defect states inside the bandgap, which are labeled by a dashed line in Fig. 3(c). In this case, the defect gap is 1.03 eV, corresponding to the energy gap between these unoccupied defect levels and the highest occupied state, i.e., the VBM of $MoSe_2$. The defect gap (1.03 eV) is comparable with the intrinsic bulk bandgap (1.43 eV) of monolayer $MoSe_2$, making it another ideal example of large defect-gap systems.

The direct *GW* results are presented in Fig. 3(d), in which the bulk bandgap is increased to 2.36 eV and the defect gap is increased to 1.78 eV. These results agree with previous *GW* calculations [71]. The GW results within the intrinsic screening approximation are shown in Fig. 3(e). The bulk band gap is 2.39 eV, and the defect gap is 1.86 eV, respectively, which agree well with the direct *GW* results. The results using the defect-patched screening approximation are shown in Fig. 3(f). As expected, the calculated bulk band gap is 2.38 eV, and the defect gap is 1.82 eV, which are slightly closer to the direct *GW* result.

**C. Single boron vacancy in monolayer *h*-BN**

In this section, we study the charge-neutral single boron vacancy ($V_B$) in monolayer *h*-BN, where a single boron is removed to form a vacancy defect as Fig. 4(a) shows. A 9 × 9 × 1 *h*-BN supercell is constructed corresponding to a 1.2% vacancy concentration. Following previous works [72], DFT calculation is performed using the local spin density approximation (LSDA) [73]. A q-grid of 2 × 2 × 1 is adopted to calculate the dielectric function and quasiparticle energies in the supercell. Around 8,000 conduction states are included in the direct *GW* calculation.

The DFT band structure is shown Fig. 4(b), and the corresponding band alignment is plotted in Fig. 4(c). There are three in-gap defect states, and all of them are unoccupied. Different from the defects discussed in Section 3 A and B, these in-gap defect states are close to the VBM of *h*-BN. As marked in Fig. 4 (c), the spin-up defect gap is 151 meV, and the spin-down defect gap is 320 meV. Thus, this is a good example of studying narrow defect-gap systems.

The result of the direct *GW* calculation is presented in Fig. 4(d), where the quasiparticle bulk bandgap of *h*-BN is increased to 7.05 eV. The spin-up defect gap is increased to 2.66 eV, and the spin-down defect gap is increased to 2.87 eV. Fig. 4 (e) shows the *GW* results within the intrinsic screening approximation. The bulk band gap is increased to 7.26 eV, the spin-up defect gap is 2.84 eV, and the spin-down defect gap is 3.13 eV. Compared to the direct *GW* results, the deviations are around 200-300 meV, which are significantly larger than those large defect-gap point defects discussed in Section 3 A and B.

This deviation is because the defect gap of the $V_B$ defect in *h*-BN is significantly smaller than the bulk bandgap. According to Eq. (3), it leads to a small transition energy and hence a sizable contribution associated defect levels. Therefore, for narrow defect-gap systems, it is essential to include the defect contributions, leading us to the defect-patched screening approximation. The defect-patched screening results are presented in Fig. 4(f). We find a spin-up defect gap of 2.74 eV and a spin-down defect gap of 2.96 eV. The deviations from the direct *GW* calculation are reduced to 80 meV and 90 meV for the spin-up and spin-down defect gaps, respectively.

This improvement can also be quantitatively explained by comparing the magnitude of the matrix elements of RPA polarizability, as plotted in Fig. 5. Total polarizability by the direct $GW$ calculation is presented in Fig. 5(a), and the defect contribution is plotted in Fig. 5(b). Comparing to those of large defect-gap systems shown in Fig. 2(a) and (c), we can clearly see that the defect contribution in the narrow defect-gap system is much larger. The matrix elements of $\chi^0_{\text{Defect}}$ is only about one order of magnitude smaller than those of directly calculated $\chi^0$ in this charge-neutral single boron vacancy. Interestingly, we notice that the improvement of the defect-patched approximation is about 200-300 meV, which is also an order of magnitude of smaller than the $GW$ self-energy corrections. Finally, Fig. 5(c) shows the validity of the defect-patched approximation, where the matrix elements of $\chi^0 - \chi^0_{\text{Intrinsic}} - \chi^0_{\text{Defect}}$ are near negligible. Therefore, for point defect systems with a defect gap significantly smaller than the bulk bandgap, it is necessary to use the defect-patched screening approximation.

### D. Negatively charged NV center in diamond

Negatively charged NV centers in diamond are discussed in this section as an example of a charged point defect in a bulk material. The NV center consists of a substitutional nitrogen atom associated with a vacancy in the neighboring lattice site as shown in Fig. 6(a). NV centers are well-known candidates for realizing quantum qubit, and their spin can be initialized and read out through spin dependent decay processes by an optical approach [22,23]. We use a 3 × 3 × 3 supercell that contains 215 atoms to model this defect system, corresponding to a 0.46% vacancy concentration. A single Γ-point sampling is employed in the supercell calculations. Around 4,000 conduction states are included in the direct $GW$ calculation.

The DFT band structure and corresponding band alignment are shown in Figs. 6(b) and 6(c), respectively. The DFT-calculated bulk bandgap of diamond is 4.27 eV. As shown in Fig. 6(c), there are three occupied spin-up defect states inside the bulk bandgap, one singly occupied with lower energy and one pair of doubly occupied states with higher energy. There are three spin-down defect states, one occupied state and two doubly degenerate unoccupied states. The spin-conserved defect gap is 1.88 eV, measured as the energy difference between the lowest unoccupied spin-down defect state and the highest occupied spin-down defect state. The spin non-conserved defect gap is 1.27 eV, measured as the minimum energy difference between unoccupied and occupied defect states without considering spin. The quasiparticle energy levels calculated by the direct $GW$ are shown in Fig. 6(d). The bulk bandgap increases to 5.14 eV. The spin-conserved gap increases to 2.81 eV, and the spin non-conserved gap increases to 2.14 eV. These results agree well with previous DFT and $GW$ calculations [47,74].

We first show the results from the intrinsic screening approximation in Fig. 6(e). The quasiparticle bulk bandgap of diamond is 5.15 eV, in good agreement with the direct $GW$ result. The spin-conserved defect gap is 2.79 eV, and the spin non-conserved defect gap is 2.36 eV. Compared to the direct $GW$ result in Fig. 6(d), there is a 220 meV deviation in the spin non-conserved gap, indicating that the screening effect from the charged NV center defect cannot be completely neglected. This also agrees with the aforementioned picture that, if the defect gap is significantly smaller than the bulk bandgap, the defect contribution to the polarizability is non-negligible.

As expected, we find that the defect-patched screening approximation substantially improves the convergence. As shown in Fig. 6(f), the spin-conserved defect gap is corrected to 2.81 eV and spin non-conserved defect gap is corrected to 2.14 eV. The deviations from the direct $GW$ results are reduced to within 10 meV. Such excellent agreements justify again the defect-patched screening approximation for small defect-gap systems.

## 4. Discussions.

### A. Point defects in the dilute-defect limit

In all results shown above, we model the point defects via the supercell approach, in which the finite-size supercells inevitably introduce artificial errors. We take the charge-neutral single boron vacancy $V_B$ in monolayer *h*-BN as an example. The bulk bandgap in a defect-containing $9 \times 9 \times 1$ *h*-BN supercell is 7.05 eV in the direct *GW* calculation as Fig. 4(d) shows, while the *GW*-calculated bandgap of pristine *h*-BN is found to be 7.33 eV. This artificial difference is from the defect-induced structural distortions and defect-defect interactions between supercells. In principle, the bulk bandgap from the supercell calculation will approach the value of pristine structure by choosing larger-size supercells. This is the dilute-defect limit, which is the typical experimental situations. However, the brute-force method to approach this limit is usually formidable for currently available computing capability.

We propose a linear interpolation approach to correct defect levels to the dilute-defect limit based on the finite-size supercell. Motivated by the idea in [74,75], we assume that the magnitude of dilute-defect limit correction for defect level is proportional to the size of defect gap and the linear dependence is the same as the bulk states vs the bulk band gap, where $\Delta_{\text{defect}} / E_g^{\text{defect}} = \Delta_{\text{bulk-gap}} / E_g^{\text{bulk-gap}}$. Then, the dilute-defect limit correction to supercell-calculated defect levels can be written as:

$$\Delta_{\text{defect}} = \frac{E_g^{\text{defect}}}{E_g^{\text{bulk-gap}}} \times \Delta_{\text{bulk-gap}} \tag{4}$$

$E_g^{\text{defect}}$ is the defect gap of the defect-containing supercell, and $E_g^{\text{bulk-gap}}$ is the bulk bandgap of the pristine structure. $\Delta_{\text{bulk-gap}}$ is the difference of the bulk bandgap between the defect supercell and pristine structure.

We have employed this correction scheme to all calculated defects, which are summarized in Table 1. Take $V_B$ in monolayer *h*-BN as an example. Prior to the dilute-defect limit correction, the energy differences of the defect gaps are 80-90 meV between the direct *GW* and defect-patched approximation. After including the dilute-defect limit correction, these differences decrease to 10-20 meV. As shown in Table 1, this dilute-defect limit correction scheme generally and substantially improved the agreements for our studied defects. It is worth mentioning that the results within the intrinsic screening approximation are also improved at this limit. This is because the density of states associated with defects becomes smaller at the dilute-defect limit, reducing their contribution to the total polarizability.

**B. Computational efficiency**

We compare the computational cost of our approaches with that of the direct calculation of the RPA polarizability. For the calculation of the single boron vacancy $V_B$ in monolayer *h*-BN, the direct calculation takes 1,609 core hours and requires 12,679 MB for the memory. As a comparison, the intrinsic screening approximation only takes 57 core hours (excluding the time spent on reciprocal-space folding, which is negligible) and only requires 338 MB for the memory, thanks to the much smaller number of **G** vectors in a unit cell (111 vectors) compared to that in a supercell (8445 vectors). These two values are 131 core hours and 354 MB for the memory, respectively, for the defect-patched screening approximation. In this example, the calculation used the stampede2 supercomputer (Intel Xeon Phi 7250) at the Texas Advanced Computing Center (TACC). $\chi^0_{\text{Defect}}$ includes significantly fewer bands (51 valence bands and 56 conduction bands) compared to 323 valence bands and about 8000 conduction bands in the direct calculation. Therefore, this approach can reduce the computational cost significantly. Moreover, the large memory requirement of direct *GW* calculations is often the prohibitive factor for routine *GW* calculations of large supercells. Both the intrinsic screening approximation and the defect-patched screening approximation can reduce the memory requirement by about two orders of magnitude, making the calculations more tractable, even on medium-sized clusters. It is worth mentioning that, in actual calculations, different architecture of the computer cluster can have different performance and wall-time hours. However, the qualitative trend in the overall saving of computational resources should be similar.

## 5. Conclusion

In this work, we accelerate the *GW* calculation of defect levels of point defects via approximations to the noninteracting RPA polarizability ($\chi^0$), which is the main bottleneck in large-scale *GW* calculations. Two levels of approximations are proposed. The first approximation is the intrinsic screening approximation, where $\chi^0$ of the supercell with point defects is replaced by that of a primitive cell via the reciprocal-space folding procedure. The second approximation is the defect-patched screening approximation, where the defect contribution to the $\chi^0$ is explicitly calculated within a small energy window, which will then be combined with the polarizability of the perfect structure to approximate that of the defect-containing supercell. These approximations substantially reduce the computational cost for RPA polarizability without sacrificing accuracy.

We apply this approach to study several typical points defects. We found that for large defect-gap systems, the intrinsic screening approximation is accurate, while for narrow defect-gap systems, it is essential to use the defect-patched screening approximation. We expect that these two approaches are generally applicable to a variety of point defect systems and speed up the search for appropriate defects for desired applications.

**FIGURES**:

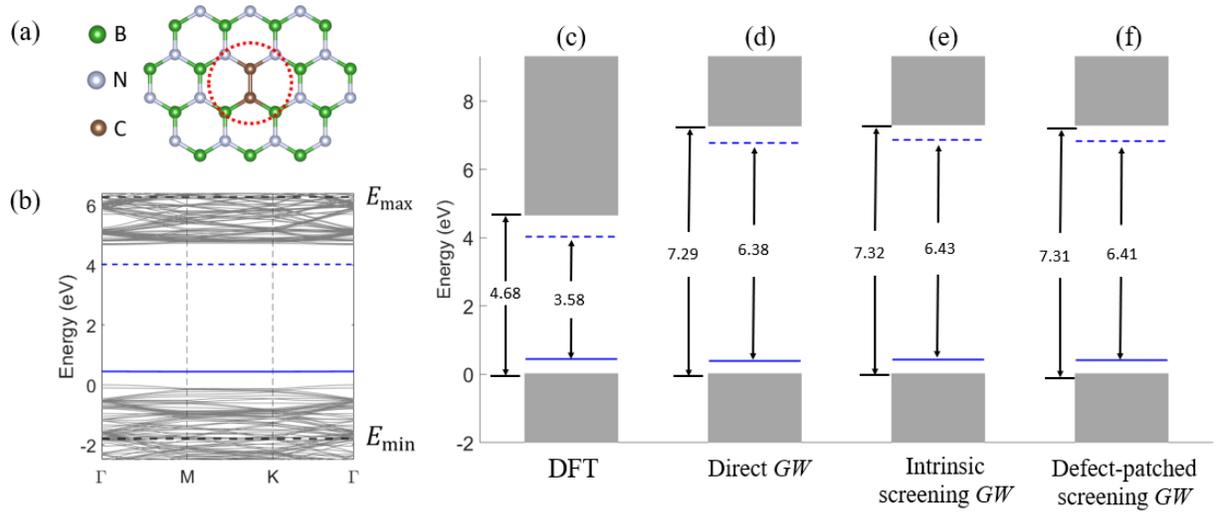

**Figure 1.** (a) Crystal structure of double carbon substitutional defects ($C_BC_N$) in monolayer *h*-BN. The defect site is circled by a red dotted line, where two carbon atoms replace a boron atom and a neighboring nitrogen atom. (b) DFT band structure for the defect system. The grey lines indicate continuum bulk bands and defect states lying inside these continuum bands. The two blue lines indicate in-gap defect states. $[E_{min}, E_{max}]$ is the energy window used in the defect-patched screening approximation. Schematic electronic structure of the defect system calculated using (c) DFT, (d) direct *GW*, (e) intrinsic screening approximation, and (f) defect-patched screening approximation. The solid (dashed) blue line represents the occupied (unoccupied) defect state. The gray boxes represent the continuum bulk bands. The VBM is set to be zero.

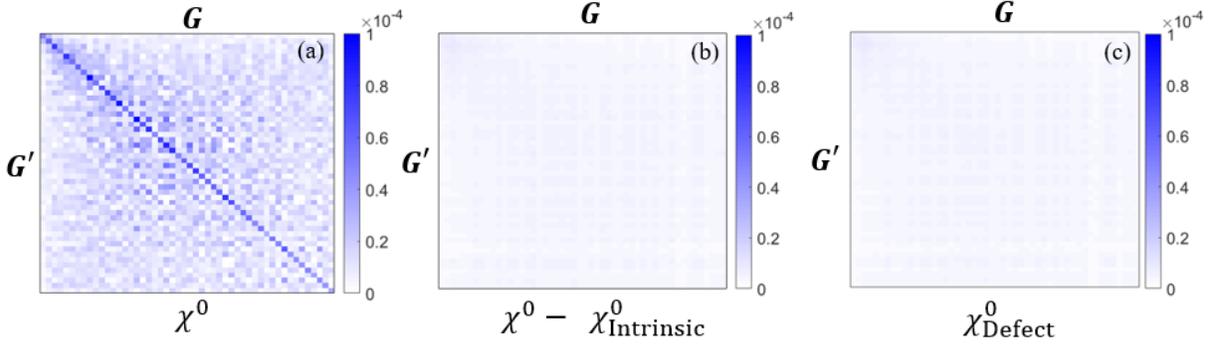

**Figure 2.** Color maps of the magnitude of matrix elements of the noninteracting RPA polarizability $\chi^0_{GG'}(q=0, \omega=0)$ for a double carbon substitutional defect $C_BC_N$ in monolayer $h$-BN. (a) Directly calculated $\chi^0$ of the defect system. (b) $\chi^0 - \chi^0_{\text{Intrinsic}}$, which is about two orders of magnitude smaller than the directly calculated $\chi^0$. (c) Defect contribution to the polarizability, $\chi^0_{\text{Defect}}$, which is calculated with Eq. (3). $\chi^0_{\text{Defect}}$ is also about two orders of magnitude smaller than the directly calculated $\chi^0$.

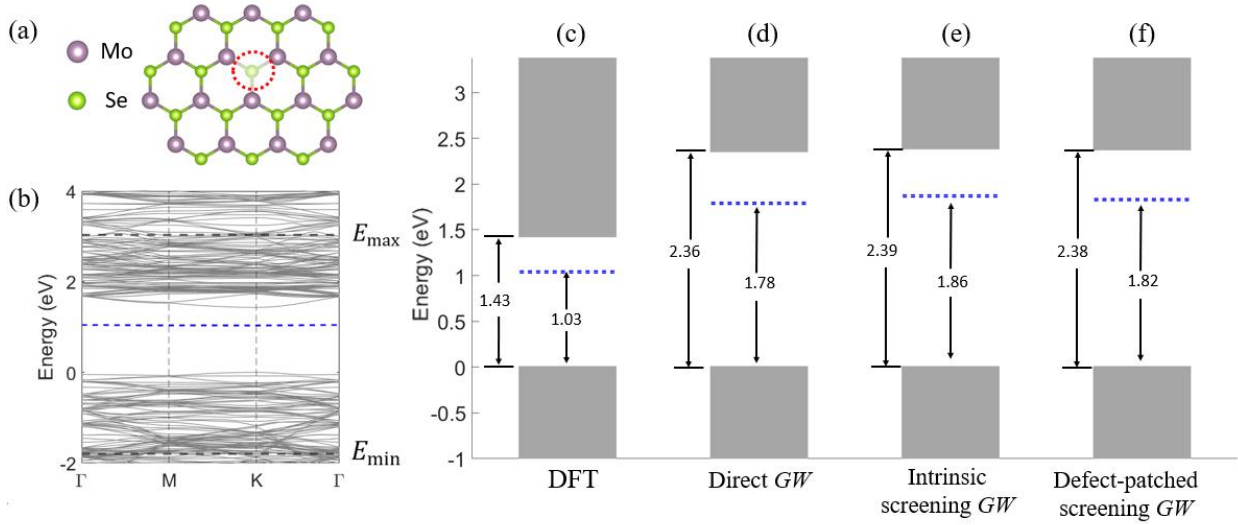

**Figure 3.** (a) Crystal structure of $V_{Se}$ in monolayer $MoSe_2$. The defect site is circled by a red dotted line. The half-transparent ball indicates the bottom Se atom, with the top Se atom removed to form a vacancy. (b) DFT band structure for the $V_{Se}$ defect system. The grey lines indicate continuum bulk bands and defect states lying inside the continuum bands. Blue lines indicate in-gap defect states, which are nearly degenerate and unoccupied. $[E_{min}, E_{max}]$ is the energy window used in the defect-patched screening calculation, where $E_{max}$ is set to be 1.6 eV above the CBM, and $E_{min}$ is set to be 1.8 eV below the VBM. Schematic electronic structure of this defect system calculated using (c) DFT; (d) direct *GW*; (e) intrinsic screening approximation; and (f) defect-patched screening approximation. The dashed line represents the nearly doubly degenerate unoccupied defect states in the gap. The gray boxes represent the continuum bulk bands. The VBM is set to be zero.

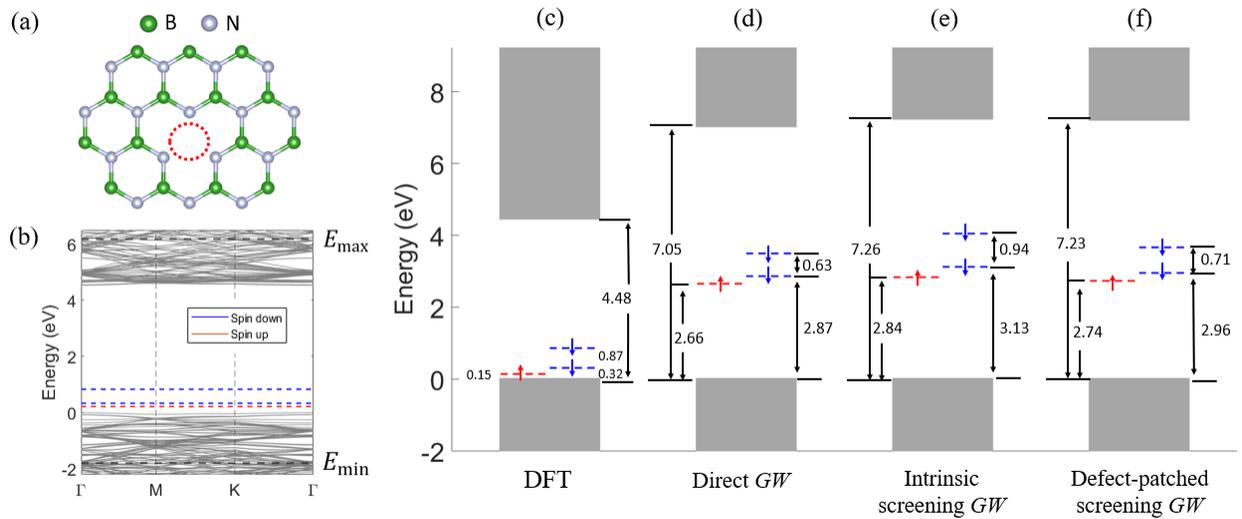

**Figure 4.** (a) Crystal structure of $V_B$ in monolayer $h$-BN. The defect site is circled by a red dotted line, where a boron atom is removed. (b) DFT band structure for the $V_B$ defect. The grey lines indicate continuum bulk bands and defect states lying inside the continuum bands. Blue and red lines indicate spin-down and spin-up defect states in the gap, respectively. $[E_{min}, E_{max}]$ is the energy window used in the defect-patched screening calculation, where $E_{max}$ is set to be 1.7 eV above the CBM and $E_{min}$ is set to be 1.8 eV lower than the VBM. Schematics of the electronic structure calculated from (c) DFT; (d) direct $GW$; (e) intrinsic screening approximation; and (f) defect-patched screening approximation. The dashed lines represent unoccupied defect states in the gap, with red (blue) color showing the spin-up (spin-down) bands The gray boxes represent the continuum bulk bands. The VBM is set to be zero.

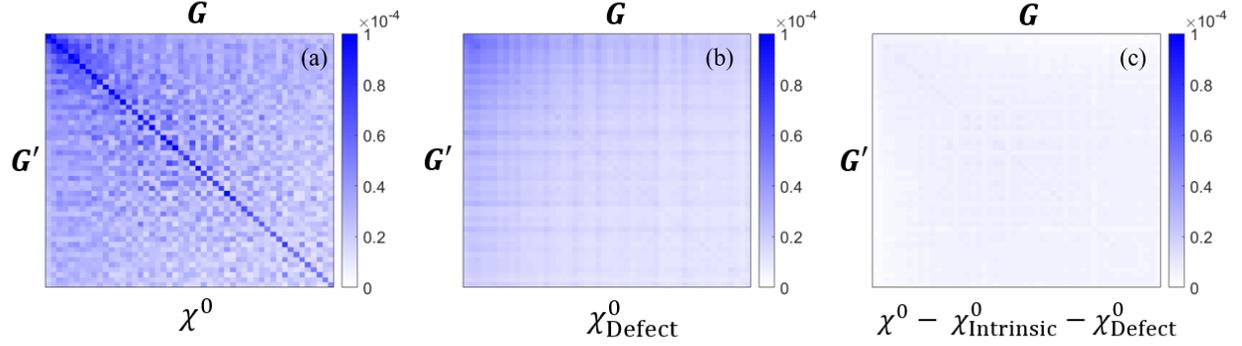

**Figure 5.** Color maps of the magnitude of matrix elements of the noninteracting RPA polarizability $\chi^0_{GG'}(q=0,\omega=0)$ for a single boron vacancy V$_B$ in monolayer *h*-BN. (a) Directly calculated $\chi^0$ of the defect system. (b) Defect contribution to the polarizability, $\chi^0_{\text{Defect}}$, which is calculated using Eq. (3). (c) Difference between the directly calculated $\chi^0$ for the defect system and the sum of $\chi^0_{\text{Defect}}$ and $\chi^0_{\text{Intrinsic}}$. This is the error in $\chi^0$ from the defect-patched screening approximation.

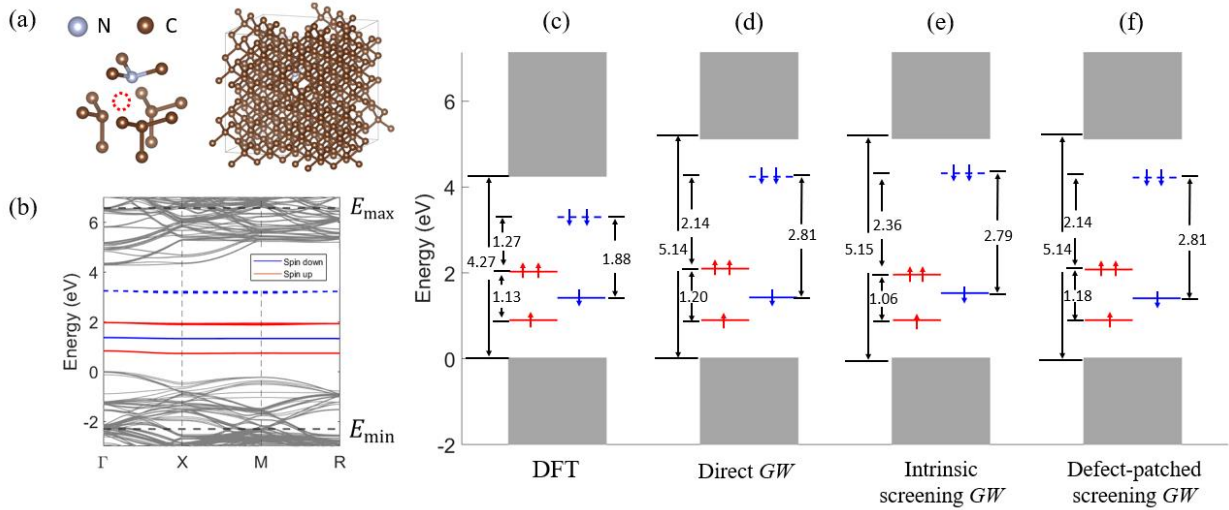

**Figure 6.** (a) Crystal structure of negatively charged NV center in diamond. The inset shows the detail of the structure of the NV center, where a carbon atom is replaced by a nitrogen atom and the neighboring carbon atom is removed, as shown by the red circle. (b) DFT band structure for the NV center defect structure. The red lines represent spin-up states, and the blue lines represent spin-down states. The grey lines indicate continuum bulk bands and defect states lying inside the continuum bands. $[E_{min}, E_{max}]$ is the energy window used in the defect-patched screening calculation, where $E_{max}$ is set to be 2.3 eV above the CBM and $E_{min}$ is set to be 2.3 eV below the VBM. Schematics of the defect electronic structure from (c) DFT; (d) direct *GW*; (e) intrinsic screening approximation; and (f) defect-patched screening approximation. The solid lines represent occupied defect states, and the dashed lines represent unoccupied defect states in the gap. The gray boxes represent the continuum bulk bands. The arrows indicate spin-up (red) or spin-down (blue) states. The VBM is set to be zero.

| Defect gap (eV) | Finite-size supercell | | | Dilute-defect limit | | |
|---|---|---|---|---|---|---|
| | Direct *GW* | Intrinsic screening *GW* | Defect-patched screening *GW* | Direct *GW* | Intrinsic screening *GW* | Defect-patched screening *GW* |
| $C_B C_N$ in monolayer *h*-BN | 6.38 | 6.43 | 6.41 | 6.42 | 6.44 | 6.43 |
| $V_{Se}$ in monolayer $MoSe_2$ | 1.78 | 1.86 | 1.82 | 1.81 | 1.87 | 1.83 |
| $V_B$ in monolayer *h*-BN — Spin-up | 2.66 | 2.84 | 2.74 | 2.77 | 2.87 | 2.78 |
| $V_B$ in monolayer *h*-BN — Spin-down | 2.87 | 3.13 | 2.96 | 2.98 | 3.16 | 3.00 |
| NV center in diamond — Spin-conserved | 2.81 | 2.79 | 2.81 | 2.82 | 2.79 | 2.82 |
| NV center in diamond — Spin-non-conserved | 2.14 | 2.36 | 2.14 | 2.14 | 2.36 | 2.14 |

**Table 1.** Defect gaps of different systems in a finite size supercell and dilute-defect limit, respectively.

## AUTHOR INFORMATION

**Corresponding Author**

*(Z.-F.L.) E-mail: zfliu@wayne.edu

†(L.Y.) E-mail: lyang@physics.wustl.edu

**Notes**

The authors declare no competing financial interest.

## ACKNOWLEDGMENT

D.L. is supported by the National Science Foundation (NSF) grant No. DMR-2124934. L.Y. is supported by NSF grant No. DMR-2118779. Z.-F.L. is supported by an NSF CAREER award, DMR-2044552. This work used Anvil at Purdue University through allocation DMR100005 from the Advanced Cyberinfrastructure Coordination Ecosystem: Services & Support (ACCESS) program, [77] which is supported by National Science Foundation grants #2138259, #2138286, #2138307, #2137603, and #2138296.